\documentclass[reprint,superscriptaddress,aps,pra]{revtex4-2}

\usepackage{amsmath,amssymb,bm}
\usepackage{graphicx}
\usepackage{braket}
\usepackage[
  colorlinks=true,
  linkcolor=blue,
  citecolor=blue,
  urlcolor=blue
]{hyperref}
\usepackage{cleveref}
\crefname{equation}{Eq.}{Eqs.}
\crefname{figure}{Fig.}{Figs.}
\crefname{section}{Sec.}{Secs.}
\usepackage[caption=false]{subfig}

\usepackage[percent]{overpic}
\usepackage{comment}

\begin{document}

\title{Geometric modulation of transition and survival intensities in non-Hermitian systems}

\author{Ryu Frieson}
\affiliation{Department of Applied Physics, Hokkaido University, Hokkaido 060-8628, Japan}
\author{Takashi Oka}
\affiliation{The Institute for Solid State Physics, The University of Tokyo, Chiba 277-8581, Japan}
\author{Hideaki Obuse}
\affiliation{Department of Applied Physics, Hokkaido University, Hokkaido 060-8628, Japan}
\affiliation{The Institute for Solid State Physics, The University of Tokyo, Chiba 277-8581, Japan}
\affiliation{Institute of Industrial Science, The University of Tokyo, Chiba 277-8574, Japan}
\affiliation{Institute for Frontier Education and Research on Semiconductors, Hokkaido University, Hokkaido 060-0808, Japan}

\date{\today}

\begin{abstract}
The time evolution of non-Hermitian systems is generally nonunitary. Dynamics governed by time-dependent non-Hermitian Hamiltonians lead to a variety of novel phenomena, one of which is state amplification or suppression induced by the complex Berry phase. Here, we extend the framework of geometric modulation to multi-level systems and show that both transition and survival intensities can be modulated. We apply our theory to the non-Hermitian Landau-Zener (LZ) problem. First, we show that, in the half-LZ problem, both the transition and survival probabilities exhibit nonreciprocity due to the complex Berry phase. In the non-Hermitian standard LZ problem, only the survival intensity is known to exhibit nonreciprocity, whereas the transition intensity does not. However, the physical origin of this nonreciprocal behavior remains unclear. In this work, we show that the nonreciprocity originates from the complex Berry phase.
\end{abstract}

\maketitle

\section{introduction}\label{sec:introduction}

Non-Hermitian quantum mechanics effectively describes open quantum systems \cite{gamow_zur_1928,feshbach_unified_1958,feshbach_unified_1962,ashida_non-hermitian_2020}. Its unique features have attracted considerable attention and have led to various applications in classical systems, including classical optical systems \cite{ashida_non-hermitian_2020,liertzer_pump-induced_2012,fleury_invisible_2015,zhang_acoustic_2021,wang_generating_2021,ghatak_observation_2020,wang_non-hermitian_2022,hofmann_reciprocal_2020,zou_observation_2021}. Experiments confirming theoretical prediction by non-Hermitian quantum mechanics have also been realized in quantum-regime systems \cite{wu_observation_2019,li_observation_2019,dogra_quantum_2021,ding_experimental_2021,liang_dynamic_2022,zhao_two-dimensional_2025}. Among them, parity-time (PT) symmetric \cite{bender_real_1998,bender_generalized_2002} systems have been actively studied \cite{bender_making_2007,guo_observation_2009,ruter_observation_2010,chong_mathcalpmathcalt-symmetry_2011,el-ganainy_non-hermitian_2018,ozdemir_paritytime_2019}. Hamiltonians protected by PT symmetry can retain real spectra, and this property is generalized to quasi-Hermiticity \cite{mostafazadeh_pseudo-hermiticity_2002-2,mostafazadeh_pseudo-hermiticity_2002,mostafazadeh_pseudo-hermiticity_2002-1}. Without exceptional points and complex eigenvalues, quasi-Hermitian systems can still exhibit nontrivial behavior, such as the non-Hermitian skin effect in the Hatano-Nelson model \cite{hatano_localization_1996,hatano_vortex_1997} under open boundary conditions.

In non-Hermitian systems, time evolution is in general nonunitary, even if the Hamiltonian is quasi-Hermitian \cite{fring_unitary_2016}. One intriguing aspect of this dynamics is the complex Berry phase \cite{berry_quantal_1984}. The non-Hermitian Berry phase \cite{garrison_complex_1988,dattoli_geometrical_1990,keck_unfolding_2003} is generally complex, and its imaginary part is gauge invariant even for noncyclic time evolution \cite{silberstein_berry_2020,singhal_measuring_2023}. In quasi-Hermitian systems, this complex Berry phase induces amplification or decay of the intensity \cite{garrison_complex_1988,dattoli_geometrical_1990,keck_unfolding_2003,silberstein_berry_2020,singhal_measuring_2023,ozawa_geometric_2025,barge_measurement_2026} even though all the eigenenergies are real. Such geometric modulation has recently been observed in a mechanical system \cite{singhal_measuring_2023} and also in a superconducting transmon circuit \cite{barge_measurement_2026}.

Another example of counterintuitive dynamics in quasi-Hermitian systems is the Landau-Zener (LZ) transition \cite{landau_zur_1932,zener_non-adiabatic_1932}. The LZ model describes nonadiabatic transitions in a two-level system. It is one of the most fundamental problems in time-dependent quantum systems and has been studied in a wide range of fields \cite{ivakhnenko_nonadiabatic_2023,oliver_mach-zehnder_2005,sillanpaa_continuous-time_2006,higuchi_light-field-driven_2017,olson_tunable_2014,faust_nonadiabatic_2012}. The Hamiltonian depends on time through a time-varying external field. In the Hermitian case, the standard LZ transition probability does not depend on the sweep direction. Hamiltonians with a certain geometric property have also been shown to exhibit nonreciprocity \cite{berry_geometric_1997}. Recently, it has been shown that nonreciprocal LZ transitions can arise from the shift vector \cite{kitamura_nonreciprocal_2020,takayoshi_nonadiabatic_2021}. In the non-Hermitian case \cite{torosov_pseudo-hermitian_2017,shen_landau-zener-stuckelberg_2019,longstaff_nonadiabatic_2019,longhi_non-bloch-band_2020,wang_adiabaticity_2022,wang_nonlinear_2023,kivela_quantum_2024,zhao_real-time_2025}, although the transition intensity remains independent of the sweep direction, the survival intensity shows nonreciprocity \cite{torosov_pseudo-hermitian_2017,shen_landau-zener-stuckelberg_2019}. These phenomena have also been observed experimentally \cite{kivela_quantum_2024,zhao_real-time_2025}. However, the origin of this nonreciprocal behavior remains unclear.

In the present work, we reveal that the nonreciprocal nature of the survival intensity originates from the complex Berry phase. To this end, we first derive a general formula for the transition and survival intensities in quasi-Hermitian systems by applying adiabatic perturbation theory (APT) \cite{rigolin_beyond_2008,de_grandi_adiabatic_2010,weinberg_adiabatic_2017,huang_direct_2025}, assuming slowly varying parameters to treat nonadiabatic effects perturbatively. We reveal that both transition and survival dynamics are governed by complex Berry phases, whose imaginary parts induce amplification or suppression of their intensities. Then, we apply our theory to the non-Hermitian LZ problems. First, we consider a half-LZ problem, where the sweeping starts from $t=0$ (not $-\infty$). We demonstrate that both transition and survival intensities exhibit nonreciprocal behavior due to the complex Berry phases. Finally, we show that, in the adiabatic limit, the results obtained in \cref{sec:apt} account for the emergence of nonreciprocity in the survival intensity of the standard non-Hermitian LZ problem. Furthermore, by considering the Berry-phase contribution in each process, we explain why the nonreciprocity appears only in the survival intensity.

This paper is organized as follows. In \cref{sec:apt}, we provide a framework to derive transition and survival intensities in quasi-Hermitian systems based on APT. \cref{sec:ex} is devoted to clarifying how the complex Berry phase affects the non-Hermitian LZ problem. By applying the general formula in \cref{sec:apt}, we first demonstrate that the complex Berry phase induces nonreciprocity in both the transition and survival intensities in the half-LZ problem. We then turn to the standard LZ problem and reveal the geometric origin of the nonreciprocal behavior. A summary and discussions are presented in \cref{sec:conclusion}.

\section{adiabatic perturbation theory for quasi-Hermitian Hamiltonians}
\label{sec:apt}
In this section, we derive the formula for the transition and survival intensities in general quasi-Hermitian systems by applying APT.

We consider a time-dependent Hamiltonian $H[\bm{\lambda(t)}]$ parameterized by an $N$-dimensional real vector $\bm{\lambda}=(\lambda_1,\,\lambda_2,\,\dots,\,\lambda_N)$ whose component $\lambda_i$ depends on time $t$. We assume that the parameters vary slowly, i.e., $\dot{\lambda}\equiv|\dot{\bm{\lambda}}|=|d\bm{\lambda} /dt|$ is sufficiently small. Consequently, nonadiabatic transitions are suppressed so that the nonadiabatic contribution can be treated perturbatively \cite{rigolin_beyond_2008,de_grandi_adiabatic_2010,weinberg_adiabatic_2017,huang_direct_2025}. The transition probability can be expanded in powers of $\dot{\lambda}^2$ or higher-order derivatives of $\bm{\lambda}$, while the survival probability is of order unity.

To formulate the above discussion, first, we introduce the instantaneous eigenproblem for the time-dependent non-Hermitian Hamiltonian $H[\bm{\lambda}(t)]$ as
\begin{align}
H[\bf{\lambda}(t)] \Ket{n,\bm{\lambda}} &= E_n (\bm{\lambda}) \Ket{n,\bm{\lambda}}, \\
\langle\Bra{n,\bm{\lambda}} H[\bf{\lambda}(t)] &=  E_n (\bm{\lambda}) \langle\Bra{n,\bm{\lambda}},
\end{align}
where $E_n (\bm{\lambda})$ is the instantaneous eigenenergy and the $\Ket{n,\bm{\lambda}}$ and $\langle\Bra{n,\bm{\lambda}}$ denote right and left instantaneous eigenstates, respectively. It is convenient to express these eigenstates satisfying the biorthonormality condition $\langle\Braket{m,\bm{\lambda} | n,\bm{\lambda}} = \delta_{mn}$ as well as the normalization condition for the right eigenstates $\Braket{n,\bm{\lambda} | n,\bm{\lambda}} = 1$.

Since we focus on the quasi-Hermitian Hamiltonian $H[\bm{\lambda}(t)]$ satisfying $\eta H[\bm{\lambda}(t)] \eta^{-1} = H[\bm{\lambda}(t)]^\dagger$ with a positive-definite operator $\eta$, at any time $t$, all the instantaneous eigenenergies $E_n(\bm{\lambda})$ are real. Moreover, since exceptional points are absent in quasi-Hermitian systems, the adiabatic theorem can be applied in the same manner as in Hermitian systems.

The time-dependent Schr\"{o}dinger equation is written as
\begin{equation}
    i\frac{\partial}{\partial t}\Ket{\psi(t)}=H\left[\bm{\lambda}(t)\right]\Ket{\psi(t)}\label{eq:TDSE},
\end{equation}
where $\Ket{\psi(t)}$ is the quantum state of the system. We expand the state $\Ket{\psi(t)}$ in the basis of right eigenstates at any time $t$ as
\begin{equation}
    \Ket{\psi(t)}=\sum_n c_n (t)\Ket{n,\bm{\lambda}}\label{eq:expansion},
\end{equation}
where $c_n(t)$ is the complex expansion coefficient for the $n$-th eigenstate. For simplicity, the state $\Ket{\psi(t)}$ is assumed to be in the ground state $\Ket{1}$ at the initial time $t=t_i$, i.e., ${c_n(t_i)=\delta_{n1}}$. The intensity in the $m$-th state at $t=t_f$ is given by
\begin{equation}
    P_m = \left| \langle\Braket{m,\bm{\lambda}(t_f)|\psi(t_f)} \right|^2 =\left| c_m (t_f)\right|^2\,.
\end{equation}
Due to the nonunitary time evolution in the non-Hermitian systems, the probability is not conserved. Hence, we refer to $P_m$ as an intensity rather than a probability. $P_1$ is the survival intensity, whereas $P_{m\neq 1}$ is the transition intensity to the $m$-th state.

By substituting \cref{eq:expansion} into \cref{eq:TDSE}, and using the biorthonormality, we derive the equation for the coefficients as
\begin{equation}
    \dot{c}_m(t) +i E_m(\bm{\lambda}(t))c_m(t) = i\sum_n \dot{\bm{\lambda}}\cdot\bm{A}_{mn}(\bm{\lambda})c_n(t)\label{eq:DE},
\end{equation}
where, $\bm{A}_{mn}$ is the non-Hermitian Berry connection defined as
\begin{equation}
    \bm{A}_{mn}(\bm{\lambda}) \equiv i\langle\Bra{m,\bm{\lambda}}\nabla_{\bm{\lambda}}\Ket{n,\bm{\lambda}}\label{eq:connection}.
\end{equation}
The Berry connection is generally complex in non-Hermitian systems. The imaginary part of its diagonal component $(m=n)$ is gauge invariant \cite{silberstein_berry_2020,singhal_measuring_2023}, whereas the off-diagonal components $(m\neq n)$ induce transitions between the $m$-th and $n$-th state.

We first consider the case $m=1$ to see how the survival intensity is affected by the non-Hermitian Berry connection. If the parameters vary adiabatically, the coefficients $c_m$ $(m\neq 1)$ for weak nonadiabatic transitions are in $O(\dot{\lambda})$. Neglecting the second-order terms in $\dot{\lambda}$, \cref{eq:DE} reduces to
\begin{equation}
    \dot{c}_1(t) \simeq -i\left[E_1(\bm{\lambda}(t)) -\dot{\bm{\lambda}}\cdot\bm{A}_{11}(\bm{\lambda})\right]c_1(t).
\end{equation}
Integrating the above equation directly yields
\begin{equation}
    c_1(t) \simeq \exp{\left[ -i\int_{t_i}^{t} E_1(\bm{\lambda}(t'))dt' + i\int_C \bm{A}_{11}\cdot d\bm{\lambda} \right]}\label{eq:c1}.
\end{equation}
Here, the first term in the exponent represents the dynamical phase governed by the Schr\"{o}dinger equation \cref{eq:TDSE}, and the second term corresponds to the Berry phase accumulated along a path $C$ in the parameter space of $\bm{\lambda}$. The dynamical phase is real, while the Berry phase is complex in quasi-Hermitian systems, in general. The imaginary part of this complex phase is gauge invariant even along an open path due to the gauge invariance of ${\mathrm{Im}\bm{A}_{11}}$. The survival intensity is therefore geometrically amplified or suppressed by the imaginary part of the non-Hermitian Berry phase, given by
\begin{equation}
    P_1 \simeq \exp{\left(-2\int_C \mathrm{Im}\bm{A}_{11}\cdot d\bm{\lambda}\right)}\label{eq:survival}.
\end{equation}

Next, we consider the transition intensity. As in the case $m=1$, we neglect second-order terms in $\dot{\lambda}$, but retain the $n=m$ term in the sum in \cref{eq:DE}. Then, \cref{eq:DE} reduces to
\begin{equation}
    \begin{split}
        \dot{c}_m(t) &\simeq - i\left[E_m(\bm{\lambda}(t)) -\dot{\bm{\lambda}}\cdot\bm{A}_{mm}(\bm{\lambda})\right]c_m(t)\\
        &\quad + i\dot{\bm{\lambda}}\cdot\bm{A}_{m1}(\bm{\lambda})c_1(t).
    \end{split}
\end{equation}
Substituting \cref{eq:c1} into the above equation, we obtain
\begin{equation}
    \begin{split}
        c_m(t) &= e^{-i\int_{t_i}^t\left( E_m-\dot{\bm{\lambda}}\cdot\bm{A}_{mm} \right)dt'}\\ &\quad \times i
        \int_{t_i}^t \dot{\bm{\lambda}}\cdot\bm{A}_{m1}e^{i\int_{t_i}^{t'}\left( \Delta E_{m1}-\dot{\bm{\lambda}}\cdot \Delta \bm{A}_{m1} \right)dt''} dt',
    \end{split}
\end{equation}
where $\Delta E_{m1}= E_m-E_1$ and $\Delta \bm{A}_{m1}= \bm{A}_{mm}-\bm{A}_{11}$. To calculate further, we assume that the energy gap is sufficiently large and the parameters vary sufficiently slowly. After integrating by parts, the leading contribution to the transition intensity comes from the boundary terms at the initial and final times. This yields
\begin{equation}
    \begin{split}
        c_m(t_f) &\simeq \frac{\dot{\bm{\lambda}}(t_f)\cdot\bm{A}_{m1}(\bm{\lambda}(t_f))}{\Delta E_{m1}(t_f)}e^{-i\int_{t_i}^{t_f} E_1 dt'}e^{i\int_C \bm{A}_{11}\cdot d\bm{\lambda}}\\
        &\quad -\frac{\dot{\bm{\lambda}}(t_i)\cdot\bm{A}_{m1}(\bm{\lambda}(t_i))}{\Delta E_{m1}(t_i)}e^{-i\int_{t_i}^{t_f} E_m dt'}e^{i\int_C \bm{A}_{mm}\cdot d\bm{\lambda}}\label{eq:cm}.
    \end{split}
\end{equation}
The transition intensity is therefore given by
\begin{widetext}
    \begin{equation}
        \begin{split}
            P_m &\simeq \left|\frac{\dot{\bm{\lambda}}(t_i) \cdot\bm{A}_{m1}(\bm{\lambda}(t_i))}{\Delta E_{m1}(t_i)}\right|^2 \exp{\left(-2\int_C \mathrm{Im}\bm{A}_{mm}\cdot d\bm{\lambda}\right)}
            + \left|\frac{\dot{\bm{\lambda}}(t_f)\cdot \bm{A}_{m1}(\bm{\lambda}(t_f))}{\Delta E_{m1}(t_f)}\right|^2 \exp{\left(-2\int_C \mathrm{Im}\bm{A}_{11}\cdot d\bm{\lambda}\right)}\\
            &\quad - 2\mathrm{Re}\left[ \frac{\dot{\bm{\lambda}}(t_i)\cdot \bm{A}_{m1}(t_i)}{\Delta E_{m1}(t_i)} \left(\frac{\dot{\bm{\lambda}}(t_f)\cdot \bm{A}_{m1}(t_f)}{\Delta E_{m1}(t_f)}\right)^{*}\exp{\left(-i\int_{t_i}^{t_f} \Delta E_{m1} dt'\right)} \exp{\left(i\int_C \Delta \bm{A}_{m1} \cdot d\bm{\lambda}\right)} \right]\label{eq:transition}.
        \end{split}
    \end{equation}
\end{widetext}
The first and second terms correspond to the transitions at the initial and final times. The third term represents the interference between these processes. If the transition occurs at the initial time, the system evolves along the excited state; if it occurs at the final time, the system evolves along the ground state. Thus, the first and second terms are weighted by the Berry phases of the excited and ground states, respectively. The contribution of the Berry phase in the interference term also appears in Hermitian systems; however, these geometric modulations are unique to non-Hermitian systems.

\section{Demonstration: LZ problems}
\label{sec:ex}
In this section, we apply the results derived in \cref{sec:apt} to the non-Hermitian LZ problem as a fundamental phenomenon of time-dependent quantum systems. The Hamiltonian is given by
\begin{equation}
    H\left[\lambda(t)\right]=
    \begin{pmatrix}
        i\gamma & \lambda-i\varepsilon \\
        \lambda+i\varepsilon & -i\gamma
    \end{pmatrix}\label{eq:LZHamiltonian},
\end{equation}
where, $\varepsilon$ and $\gamma$ are real fixed parameters, and $\varepsilon$ characterized the energy gap, while the non-hermitian terms $\pm i\gamma$ correspond to gain and loss. The sweep parameter $\lambda$ represents a time-dependent external field defined by $\lambda =-Ft$. The parameter $\lambda$ varies with the sweep velocity $F$, with the sign of $F$ determining the sweep direction. This non-Hermitian Hamiltonian has been introduced to study the non-Hermitian LZ problem \cite{shen_landau-zener-stuckelberg_2019} and possesses quasi-Hermiticity with
\begin{equation}
    \eta = \frac{1}{\sqrt{\varepsilon^2-\gamma^2}}
    \begin{pmatrix}
        \varepsilon & -\gamma \\
        -\gamma & \varepsilon
    \end{pmatrix}\;.
\end{equation}
In the case of the standard LZ problem, where the ground state is swept from $t_i=-\infty$ to $t_f=\infty$, the transition and survival intensities of \cref{eq:LZHamiltonian} are analytically derived as \cite{torosov_pseudo-hermitian_2017,shen_landau-zener-stuckelberg_2019}
\begin{align}
    P_{+}(F) &= \exp{\left( -\pi\frac{\varepsilon^2-\gamma^2}{\left| F \right|} \right)}\label{eq:LZtrans},\\
    P_{-}(F) &= \left(\frac{\varepsilon+\gamma}{\varepsilon-\gamma}\right)^{\mathrm{sgn}F}\left(1-P_{+}(F)\right)\label{eq:LZsurv}.
\end{align}
using the parabolic cylinder function \cite{abramowitz_handbook_1968}. Several remarks regarding \cref{eq:LZtrans,eq:LZsurv} are in order: i) The total intensity $P_+ (F)+P_{-}(F)$ is not unity due to non-Hermiticity, ii) when $\gamma=0$ \cref{eq:LZtrans,eq:LZsurv} agree with the formula in the Hermitian case, iii) $P_+(F)$ is symmetric about $F$, while $P_{-}(F)$ is not. The third property means that, remarkably, the survival intensity exhibits nonreciprocity with respect to the sweep direction in contrast to the transition intensity. More specifically, the prefactor of the survival intensity becomes $(\varepsilon+\gamma)/(\varepsilon-\gamma)$ for $F>0$ and $(\varepsilon-\gamma)/(\varepsilon+\gamma)$ for $F<0$. On the other hand, the transition intensity does not show the nonreciprocity. However, the physical origin of these behaviors has not yet been understood.

Hereafter, we explain that this phenomenon originates from the complex Berry phase based on the results in \cref{sec:apt}. However, since the LZ problem is a completely nonperturbative phenomenon, nonadiabatic transitions at finite $F$ cannot be described within APT. Therefore, we first focus on the half-LZ problem in \cref{fig:HLZ_image} where the dominant transition occurs at the initial time. We derive the transition and survival intensities of the half-LZ problem by applying \cref{eq:transition,eq:survival}. Then, we turn to the standard LZ problem. The survival intensity in the adiabatic limit is derived using \cref{eq:survival}. Finally, we explain why only the survival intensity exhibits nonreciprocity, whereas the transition intensity remains reciprocal.
\begin{figure}[t]
    \stepcounter{figure}
    \centering
    \begin{minipage}[b]{0.49\linewidth}
        \centering
        \begin{overpic}[width=\linewidth]{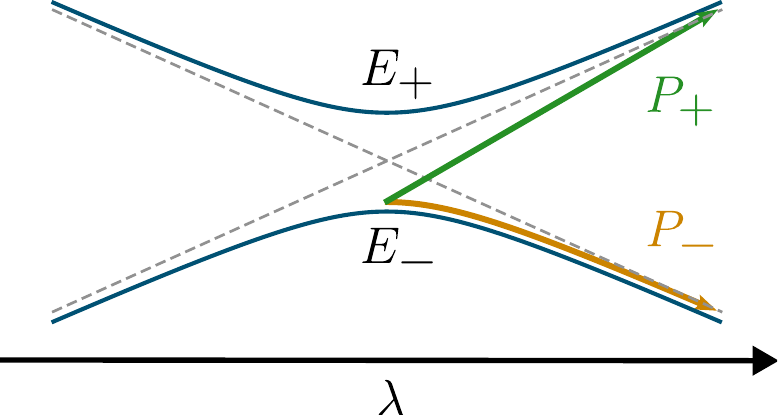}
            \put(2, 56){\large (a)}
        \end{overpic}
        \refstepcounter{subfigure}
        \label{fig:HLZ_image}
    \end{minipage}
    \hfill
    \begin{minipage}[b]{0.49\linewidth}
        \centering
        \begin{overpic}[width=\linewidth]{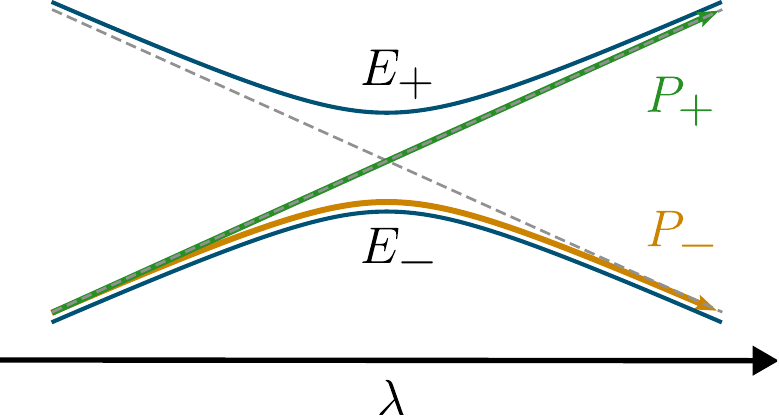}
            \put(2, 56){\large (b)}
        \end{overpic}
        \refstepcounter{subfigure}
        \label{fig:LZ_image}
    \end{minipage}
    \addtocounter{figure}{-1}
    \caption{Schematics of (a) the half-LZ problem starting at $t_i=0$ and (b) the standard LZ problem starting at $t_i=-\infty$ with $F<0$. In both cases, the system is prepared in the ground state at $t_i$ and swept to $t_f=\infty$.}
    \label{fig:schematic}
\end{figure}

\subsection{Half-Landau-Zener problem}
\label{sec:half}
In this subsection, we derive the transition and the survival intensities for the half-LZ problem.

To begin with, we derive the non-Hermitian Berry connection for the Hamiltonian in \cref{eq:LZHamiltonian}. In the parameter space of $\lambda$, the geometric property of the path is determined by $F$. The right and left instantaneous eigenstates, $\ket{\pm}$ and $\langle\bra{\pm}$, respectively, of this system associated with the instantaneous eigenenergies $E_{\pm}=\pm E$, $E=\sqrt{\lambda^2+\varepsilon^2-\gamma^2}$ are given by
\begin{subequations}
    \begin{align}
        \Ket{\pm} &= \frac{1}{\sqrt{2(\lambda^2+\varepsilon^2)}}
        \begin{pmatrix}
            \lambda -i\varepsilon \\
            \pm E-i\gamma
        \end{pmatrix}\label{eq:LZright}\,,\\
        \langle\Bra{\pm} &= \frac{\sqrt{\lambda^2+\varepsilon^2}}{\sqrt{2}(E^2\mp i\gamma E)}
        \begin{pmatrix}
            \lambda +i\varepsilon & \pm E -i\gamma
        \end{pmatrix}\label{eq:LZleft}\,.
    \end{align}
\end{subequations}
These eigenstates are constructed to satisfy both the biorthonormality and normalization conditions for the right eigenstates. To ensure that the eigenenergies are real, we require that $|\varepsilon|>|\gamma|$. Substituting \cref{eq:LZright,eq:LZleft} into \cref{eq:connection}, the non-Hermitian Berry connections are derived as
\begin{align}
    \mathrm{Im}A_{\pm\pm} &= \frac{\lambda}{2E^2}-\frac{\lambda}{2(\lambda^2+\varepsilon^2)} \mp\frac{\varepsilon\gamma}{2E(\lambda^2+\varepsilon^2)}\label{eq:diag}\,,\\
    A_{\pm\mp} &= \frac{\lambda+i\varepsilon}{2E\left(E\mp i\gamma\right)}-\frac{\lambda}{2E^2}\label{eq:offdiag}.
\end{align}

In the half-LZ problem, the system with the initial state $\ket{-}$ is swept by $\lambda$ from $t_i=0$ to $t_f=\infty$ (\cref{fig:HLZ_image}). The APT is applicable when the energy gap is sufficiently large compared to the sweep rate, i.e., ${|(\varepsilon^2-\gamma^2)/F|\gg 1}$. At the final time $t_f=\infty$, transitions do not occur as the energy gap becomes infinite. In contrast, transitions may occur predominantly when the energy gap is minimum, that is, at $t=t_i=0$. After the transition, the state evolves along the instantaneous excited state $\ket{+}$ and acquires the corresponding complex Berry phase. By substituting \cref{eq:diag,eq:offdiag} into \cref{eq:transition}, the transition intensity is obtained as
\begin{align}
    P_{+}(F) &\simeq \frac{F^2}{16\left(\varepsilon^2-\gamma^2\right)^2}\exp{\left(-2\int_0^{-\mathrm{sgn}(F)\infty} \mathrm{Im}A_{++}d\lambda\right)}\notag\\
    &= \frac{\varepsilon-\gamma\mathrm{sgn}F}{\varepsilon}\frac{F^2}{16\left(\varepsilon^2-\gamma^2\right)^2}\label{eq:HLZtrans}.
\end{align}
For the survival process, the system remains in the instantaneous ground state and acquires the corresponding Berry phase. From \cref{eq:survival,eq:diag}, the survival intensity is obtained as
\begin{equation}
    \begin{split}
        P_{-}(F) &\simeq \exp{\left(-2\int_0^{-\mathrm{sgn}(F)\infty} \mathrm{Im}A_{--}d\lambda\right)}\\
        &\quad = \frac{\varepsilon+\gamma\mathrm{sgn}F}{\varepsilon}\label{eq:HLZsurv}.
    \end{split}
\end{equation}
(See the appendix for details of the calculation.) Both expressions for the transition and survival intensities depend on $\mathrm{sgn}F$, which implies that they exhibit nonreciprocity. This contribution arises since the direction of the integration path for the Berry connection changes with $\mathrm{sgn}F$.

To confirm the above analytical result, we numerically calculated the dependence on the sweep velocity $F$ of the transition and survival intensities as shown in \cref{fig:HLZ_intensity}. We fixed $\varepsilon=1$ and $\gamma=0.5$. As shown in the inset, the numerical result of the transition intensity $P_{+}(F)$ (dots) is asymmetric for $F$ and agrees well with \cref{eq:HLZtrans} (black line) near $F=0$. We also see that the numerical result of the survival intensity $P_{-}(F)$ (dash-dotted line) discontinuously changes around $F=0$, meaning strong nonreciprocity. Remarkably, the asymptotic values of the survival intensity at $F=0$ from the positive and negative $F$ obey those predicted by \cref{eq:HLZsurv} (dashed line). Thereby, we confirmed that \cref{eq:HLZtrans,eq:HLZsurv} derived from APT are valid near $F=0$ and the imaginary part of the complex Berry phase dominates the quantum dynamics in the non-Hermitian system. We also emphasize that although APT results are valid only in the vicinity of $F=0$, this nonreciprocal behavior persists even in larger $F$, i.e., the strongly nonadiabatic regime, as confirmed in \cref{fig:HLZ_intensity}. This suggests that the complex Berry phase effects may survive even in the nonadiabatic regime.
\begin{figure}[t]
    \stepcounter{figure}
    \centering
    \begin{minipage}[b]{0.9\linewidth}
        \centering
        \begin{overpic}[width=\linewidth]{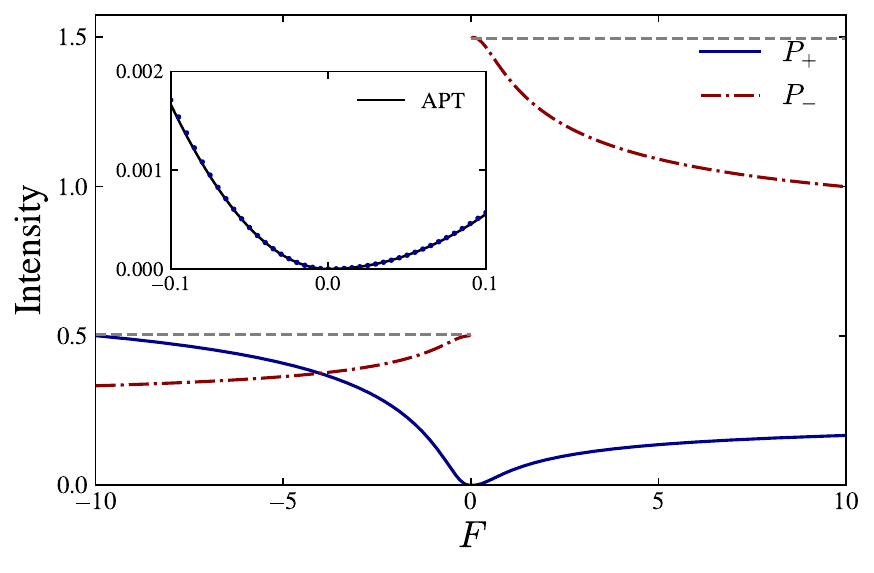}
            \put(12,66){\large (a)}
        \end{overpic}
        \refstepcounter{subfigure}
        \label{fig:HLZ_intensity}
    \end{minipage}
    
    \begin{minipage}[b]{0.9\linewidth}
        \centering
        \begin{overpic}[width=\linewidth]{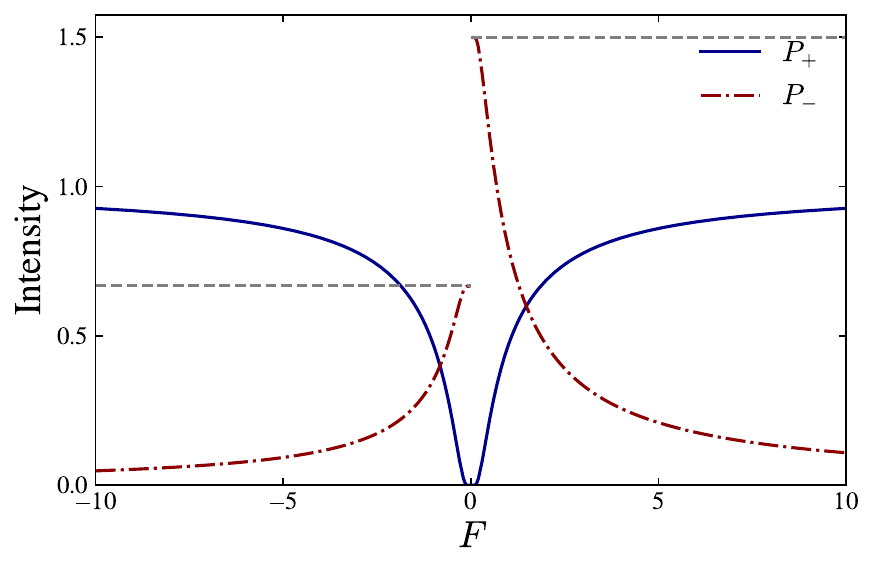}
            \put(12,66){\large (b)}
        \end{overpic}
        \refstepcounter{subfigure}
        \label{fig:LZ_intensity}
    \end{minipage}
    \addtocounter{figure}{-1}
    \caption{The sweep velocity dependence of the transition and survival intensities of (a) the half-LZ with $\varepsilon=1$ and $\gamma=0.5$ and (b) the standard LZ problems with $\varepsilon=0.5$ and $\gamma=0.1$. The numerical results of the transition and survival intensities are shown by solid blue and dash-dotted red lines, respectively. The horizontal dashed lines show the values predicted by \cref{eq:HLZsurv} for (a) and by \cref{eq:LZsurvlimit} for (b). The inset in (a) shows the transition intensity near $F=0$: numerical results (dots) and analytical results (solid black lines) obtained from APT using \cref{eq:HLZtrans}. In the numerical calculation, we set $t_i=-300$ and $t_f=300$ for (a) and $t_i=0$ and $t_f=300$ for (b).}
    \label{fig:intensity}
\end{figure}

\subsection{Origin of the nonreciprocity in the Landau-Zener problem}
\label{sec:LZ}
As seen in the half-LZ case, the nonreciprocity is governed by the imaginary part of the Berry phase. Therefore, we expect that intensities of the standard LZ problem \cref{eq:LZtrans,eq:LZsurv} are also affected by the imaginary part of the Berry phase. Here, we first derive the survival intensity \cref{eq:LZsurv} in the adiabatic limit using the result in \cref{sec:apt}. Then, we make clear why nonreciprocity does not appear in the transition intensity \cref{eq:LZtrans} by evaluating the accumulated imaginary geometric phase in both the transition and survival processes.

\subsubsection{Survival intensity}\label{sec:survival}
As mentioned above, since the standard LZ transition is intrinsically nonperturbative, the perturbative results of \cref{sec:apt} cannot be directly applied to it. In APT, intensities are expressed as a power series in the adiabatic parameter $F$. However, in the standard LZ problem, the LZ formulas \cref{eq:LZtrans,eq:LZsurv} cannot be expanded as a power series in $F$ around $F=0$ due to an essential singularity at this point. Indeed, the transition intensity calculated using \cref{eq:transition} is zero even for $F\neq 0$. Thus, the intensities for $F\neq 0$ cannot be obtained within APT. Nevertheless, the adiabatic limit $F\rightarrow 0$ can still be analyzed using the results in \cref{sec:apt}, since the nonadiabatic transitions do not occur and the system remains in the adiabatic state.

Here, we derive the survival intensity in the adiabatic limit $F\rightarrow 0$. In the survival process, the system remains in the ground state throughout the evolution and acquires the corresponding geometric phase. As we saw in the derivation of \cref{eq:HLZtrans,eq:HLZsurv}, $\mathrm{sgn}F$ is responsible for the change of the integration path for the Berry connection. In the standard LZ problem, the integration interval $t\in (-\infty,\,\infty)$ is mapped to $\infty\rightarrow-\infty$ for $F>0$ and to $-\infty\rightarrow\infty$ for $F<0$. Therefore, in the adiabatic limit, the survival intensity is obtained as
\begin{equation}
    \begin{split}
        P_{-}(F) &\simeq \exp{\left(-2\mathrm{sgn}F \int_{\infty}^{-\infty} \mathrm{Im}A_{--} d\lambda \right)}\\
        &= \left(\frac{\varepsilon+\gamma}{\varepsilon-\gamma}\right)^{\mathrm{sgn}F}\label{eq:LZsurvlimit},
    \end{split}
\end{equation}
using \cref{eq:survival,eq:diag}. This result agrees with that obtained by taking the $F\rightarrow 0$ limit in \cref{eq:LZsurv}. The dependence of this expression on $\mathrm{sgn}F$ implies nonreciprocal behavior with respect to the sweep direction. Although the present result is restricted to the adiabatic limit, the nonreciprocity is most prominent in this regime and persists even for larger $F$, as shown in \cref{fig:LZ_intensity}. This suggests that the complex Berry phase makes a robust contribution even in the nonadiabatic regime.

\subsubsection{Transition intensity}\label{sec:transition}
In \cref{sec:transition}, we discussed the origin of nonreciprocity of the survival intensity in the non-Hermitian LZ problem. However, this does not explain why only the survival intensity depends on the sweep direction, whereas the transition intensity remains reciprocal. To clarify this point, we show that the imaginary geometric phases acquired during the transition and survival processes determine whether nonreciprocity emerges in the corresponding intensities.

Nonadiabatic transitions are localized around $t=\lambda=0$, where the instantaneous energy gap is minimal. In particular, we assume the transition occurs at this point. During the transition, the system remains in the ground state until the transition at $t=0$, then evolves in the excited state. Thus, before the transition, the system acquires the Berry phase associated with the ground state:
\begin{equation}
    \exp{\left( -2\int_{\mathrm{sgn}(F)\infty}^0 \mathrm{Im}A_{--}d\lambda \right)}=\frac{\varepsilon}{\varepsilon-\gamma\mathrm{sgn}F}\label{eq:Berryto0},
\end{equation}
which is obtained from \cref{eq:survival,eq:diag}. The integration interval $t\in(-\infty,0]$ is mapped to the different intervals in the parameter space depending on the sign of $F$. On the other hand, after the transition, the system acquires the Berry phase associated with the excited state. For the interval $t\in[0,\infty)$, the corresponding path in the parameter space is likewise determined by the sign of $F$. Using \cref{eq:survival,eq:diag}, we obtain
\begin{equation}
    \exp{\left( -2\int_{\mathrm{sgn}(F)\infty}^0 \mathrm{Im}A_{++}d\lambda \right)}=\frac{\varepsilon-\gamma\mathrm{sgn}F}{\varepsilon}\label{eq:Berryfrom0}.
\end{equation}

Now, we combine these contributions from the Berry phases \cref{eq:Berryto0,eq:Berryfrom0} to obtain the amplification/suppression factor for the entire process of the standard LZ problem. In the transition process, these imaginary geometric phases acquired before and after the transition cancel out. As a result, the expression for the transition intensity \cref{eq:LZtrans} is independent of $\mathrm{sgn}F$, and thus it is reciprocal.

\section{Conclusion}
\label{sec:conclusion}
In this study, we showed that the imaginary part of the non-Hermitian Berry phase can induce intensity amplification even in multi-level systems. We derived a formula describing the geometric amplification or suppression of transition and survival intensities in general quasi-Hermitian systems. We applied our result to the non-Hermitian LZ model. In the half-LZ problem, we found that both the transition and survival intensities exhibit nonreciprocity due to the complex Berry phase. We further revealed that the nonreciprocity in the non-Hermitian LZ problem arises from these geometric phases.

It is worth discussing possible extensions of the present work. In this paper, our analysis has been restricted to PT-symmetric systems. However, many characteristic features of non-Hermitian systems, such as nonreciprocity of the LZ problem, become most prominent in the vicinity of exceptional points. Therefore, it is desirable to extend our framework to more general non-Hermitian systems with exceptional points. Such an extension may also be useful for studying geometric modulation effects in the nonadiabatic regime, including those associated with the Aharonov-Anandan phase \cite{aharonov_phase_1987}.

\begin{acknowledgments}
We thank K. Kobayashi, T. Ozawa, K. Sasaki, and H. Schomerus for their helpful discussions.
This work was supported by JSPS KAKENHI (Grants Nos. JP23K22411, JP24K00545, and JP26K00624).
H.O. was supported by the JST PRESTO Grant No. JPMJPR2454. T.O. acknowledges support from JSPS KAKENHI (Nos. JP23H04865, 26K00634, 26K00698, and 26K00662), MEXT, Japan.
\end{acknowledgments}

\appendix

\section{Derivation of \cref{eq:HLZtrans,eq:HLZsurv,eq:LZsurvlimit,eq:Berryto0,eq:Berryfrom0}}

In this appendix, we provide details of the derivation of \cref{eq:HLZtrans,eq:HLZsurv,eq:LZsurvlimit,eq:Berryto0,eq:Berryfrom0}. The complex Berry-phase contribution is evaluated by integrating the complex Berry connection in \cref{eq:diag} over the parameter $\lambda$. The integration of \cref{eq:diag} is calculated as
\begin{align}
    &2\int \mathrm{Im}A_{\pm\pm}d\lambda \notag\\
    &\quad = \int \left(\frac{\lambda}{E^2}-\frac{\lambda}{(\lambda^2+\varepsilon^2)} \mp\frac{\varepsilon\gamma}{E(\lambda^2+\varepsilon^2)}\right)d\lambda\notag\\
    &\quad =\frac{1}{2}\left( \ln{E^2}-\ln{\left( \lambda^2+\varepsilon^2 \right)} \right) \mp\frac{1}{2}\ln{\frac{\varepsilon E+\gamma\lambda}{\varepsilon E-\gamma\lambda}}\notag\\
    &\quad = \frac{1}{2}\ln{\left[\frac{\lambda^2+\varepsilon^2-\gamma^2}{\lambda^2+\varepsilon^2}\cdot\frac{\varepsilon E\mp\gamma\lambda}{\varepsilon E\pm\gamma\lambda}\right]}\label{eq:appconnection}
\end{align}
The third integral is evaluated using the formula \cite{abramowitz_handbook_1968}
\begin{equation}
    \begin{split}
        &\int \frac{dx}{(ax^2+b)\sqrt{(cx^2+d)}}\\
        &=\frac{1}{2\sqrt{b(bc-ad)}}\ln{\frac{\sqrt{b(cx^2+d)}+x\sqrt{bc-ad})}{\sqrt{b(cx^2+d)}-x\sqrt{bc-ad}}}
    \end{split}
\end{equation}
for $bc>ad$.
Evaluating \cref{eq:appconnection} at $\lambda=0$, we obtain
\begin{equation}
    \begin{split}
        &\left. \frac{1}{2}\ln{\left[\frac{\lambda^2+\varepsilon^2-\gamma^2}{\lambda^2+\varepsilon^2}\cdot\frac{\varepsilon  E\mp\gamma\lambda}{\varepsilon E\pm\gamma\lambda}\right]}\right|_{\lambda=0}\\ &\quad =\ln{\frac{\sqrt{\varepsilon^2-\gamma^2}}{\varepsilon}},
    \end{split}
\end{equation}
while taking the limit $\lambda\rightarrow\infty$ and $\lambda\rightarrow -\infty$, we obtain
\begin{equation}
    \begin{split}
        &\left. \frac{1}{2}\ln{\left[\frac{\lambda^2+\varepsilon^2-\gamma^2}{\lambda^2+\varepsilon^2}\cdot\frac{\varepsilon  E\mp\gamma\lambda}{\varepsilon E\pm\gamma\lambda}\right]}\right|_{\lambda\rightarrow\infty}\\
        &\quad = \ln{\sqrt{\frac{\varepsilon\mp\gamma}{\varepsilon\pm\gamma}}},
    \end{split}
\end{equation}
and
\begin{equation}
    \begin{split}
        &\left. \frac{1}{2}\ln{\left[\frac{\lambda^2+\varepsilon^2-\gamma^2}{\lambda^2+\varepsilon^2}\cdot\frac{\varepsilon  E\mp\gamma\lambda}{\varepsilon E\pm\gamma\lambda}\right]}\right|_{\lambda\rightarrow -\infty}\\
        &\quad = \ln{\sqrt{\frac{\varepsilon\pm\gamma}{\varepsilon\mp\gamma}}}\;,
    \end{split}
\end{equation}
respectively. Thus, integrals in\cref{eq:HLZtrans,eq:HLZsurv,eq:Berryfrom0} are calculated as
\begin{align}
    &\exp{\left(-2\int_0^{-\mathrm{sgn}(F)\infty} \mathrm{Im}A_{\pm\pm}d\lambda\right)}\nonumber\\
        &= \exp{\left[-\left( \mathrm{sgn}F\ln{\sqrt{\frac{\varepsilon\pm\gamma}{\varepsilon\mp\gamma}}} -\ln{\frac{\sqrt{\varepsilon^2-\gamma^2}}{\varepsilon}} \right)\right]}\nonumber\\
        &= \exp{\left( \ln{\frac{\varepsilon\mp\gamma\mathrm{sgn}F}{\varepsilon}} \right)}\nonumber\\
        &= \frac{\varepsilon\mp\gamma\mathrm{sgn}F}{\varepsilon}\;.
\end{align}
The integral in \cref{eq:LZsurvlimit} is calculated as
\begin{align}
    &\exp{\left(-2\mathrm{sgn}F \int_{\infty}^{-\infty} \mathrm{Im}A_{--} d\lambda \right)}\nonumber\\
        &= \exp{\left[-2\mathrm{sgn}F\left( \ln{\sqrt{\frac{\varepsilon -\gamma}{\varepsilon +\gamma}}}-\ln{\sqrt{\frac{\varepsilon +\gamma}{\varepsilon -\gamma}}} \right)\right]}\nonumber\\
        &= \exp{\left( \mathrm{sgn}F \ln{\frac{\varepsilon+\gamma}{\varepsilon-\gamma}} \right)}\nonumber\\
        &= \left(\frac{\varepsilon+\gamma}{\varepsilon-\gamma}\right)^{\mathrm{sgn}F}\;.
\end{align}
The integral in \cref{eq:Berryto0} is also calculated as
\begin{align}
    &\exp{\left(-2\int_{\mathrm{sgn}F \infty}^0 \mathrm{Im}A_{--}d\lambda\right)}\nonumber\\
        &= \exp{\left[-\left( \ln{\frac{\sqrt{\varepsilon^2-\gamma^2}}{\varepsilon}}-\mathrm{sgn}F \ln{\sqrt{\frac{\varepsilon +\gamma}{\varepsilon -\gamma}}} \right)\right]}\nonumber\\
        &= \exp{\left( \ln{\frac{\varepsilon}{\varepsilon-\gamma\mathrm{sgn}F}} \right)}\nonumber\\
        &= \frac{\varepsilon}{\varepsilon-\gamma\mathrm{sgn}F}\;.
\end{align}

\bibliography{papers}

\end{document}